\documentclass[aps,prl,twocolumn,showpacs]{revtex4}

\usepackage{graphicx}
\usepackage{latexsym}
\usepackage{bm}
\usepackage{amsmath}

\usepackage{color}

\newcommand{\comment}[1]{}

\newcommand{\lsim}{\mbox{\raisebox{-0.6ex}{$\stackrel{<}{\sim}$}}\:}

\begin{document}

\title{
QCD Plasma Instabilities and Isotropization
}

\author{Adrian Dumitru}
\author{Yasushi Nara}
\affiliation{%
Institut f\"ur Theoretische Physik,
Johann Wolfgang Goethe Universit\"at,
Max von Laue Str.\ 1,
D-60438 Frankfurt am Main, Germany
}

\date{\today}
 
\begin{abstract}
We solve the coupled Wong Yang-Mills equations for both $U(1)$ and
$SU(2)$ gauge groups and anisotropic particle momentum distributions
numerically on a lattice. For weak fields with initial energy density
much smaller than that of the particles we confirm the existence of
plasma instabilities and of exponential growth of the fields which has
been discussed previously. Also, the $SU(2)$ case is
qualitatively similar to $U(1)$, and we do find significant
``abelianization'' of the non-Abelian fields during the period of
exponential growth. However, the effect nearly disappears when the
fields are strong. This is because of the very rapid isotropization
of the particle momenta by deflection in a strong
field on time scales comparable to that for the development of
Yang-Mills instabilities. This mechanism for isotropization may lead
to smaller entropy
increase than collisions and multiplication of hard gluons,
which is interesting for the phenomenology of high-energy
heavy-ion collisions.

\end{abstract}


\maketitle


High-energy heavy-ion collisions release a large amount of partons
from the wavefunctions of the colliding nuclei. Partons
with large transverse momenta originate from high-$Q^2$ hard
interactions which can be computed from perturbative QCD~\cite{pQCD}.
On the other hand, partons with ``small'' transverse momenta on the
order of the so-called saturation momentum $Q_s$ (given by
the square root of the total color charge density
per unit rapidity and unit area
in the incoming nuclei) are much more abundant
if $Q_s\gg\Lambda_{\rm QCD}$ and are better viewed as a classical
non-Abelian field~\cite{MV}.

If the presence of the soft classical field is neglected, which amounts
to assuming that $Q_s\sim\Lambda_{\rm QCD}$, the time-evolution of the
hard partons after they come on-shell can be studied by means of the
Boltzmann equation with a collision kernel, which is the so-called
parton-cascade approach~\cite{GM,part_casc}. The collision kernel could
be truncated at the level of elastic binary collisions (perhaps
with a summation of time-like and space-like parton showers in the leading
logarithmic approximation~\cite{GM}); recently, an attempt to
fully include $2\leftrightarrow 3$ processes beyond the relaxation
time and leading-logarithmic approximations has also been
made~\cite{Xu:2004mz}.

On the other hand, for large nuclei and at high energies the
saturation scale $Q_s$ is expected to grow much larger than
$\Lambda_{\rm QCD}$~\cite{MV,Mueller} and so the presence of the
classical field can no longer be neglected. The ``bottom-up
scenario''~\cite{BMSS} generalizes the parton cascade description of
the time-evolution after the collision to include the
soft classical modes, too.
Soft gluon radiation is found to be the dominant process leading to
equilibration~\cite{BMSS,Xu:2004mz,Shin} (see also papers by Wong
in~\cite{part_casc}).

Recently, it has been argued that collective processes due to the soft
gauge field should be taken into account. Specifically, QCD plasma
instabilities may develop due to anisotropic distributions of released
hard partons~\cite{Randrup} and modify the ``bottom-up scenario''
significantly~\cite{ALM}.  The hard loop effective action for
anisotropic hard modes was formulated in~\cite{HTL} and unstable soft
modes were analyzed in~\cite{HTL_unst_mode}.  Numerical studies of its
static limit~\cite{Arnold:2004ih} revealed the interesting tendency of
the non-Abelian gauge fields to ``abelianize'' during the stage of
instability in the sense that locally commutators become much smaller
than the fields themselves (see below). The ``abelianization'' has
also been seen in solutions of the full non-linear hard loop effective
action~\cite{Rebhan:2004ur}.  It is argued that, because of
abelianization, non-Abelian effects should not cause
instabilities to saturate; rather, similarly to the Abelian case, the fields
should continue to grow until their energy density becomes comparable
to that of the hard modes~\cite{Arnold:2004ih,Rebhan:2004ur,Arnold:2004ti},
i.e.\ until the growing fields begin to have a significant effect on
the dynamics of the particles.

It is interesting to note the following difference between
isotropization by propagation of particles in a strong random field
versus that via scattering and gluon multiplication. Namely, in the
absence of a collision kernel the entropy of any specific initial
condition is conserved, while the standard parton cascade approach
produces additional entropy~\cite{part_casc,PCentropy}. An ensemble
average over sufficiently random initial field configurations can
nevertheless increase the entropy of the soft modes by a moderate
(logarithmic)
amount; this follows from the equivalence of the averaged classical field
description to a Boltzmann equation to leading and subleading orders
in the occupation number~\cite{MuellerSon}.

In heavy-ion collisions, it might not be necessary to achieve
``true'' thermalization in the sense of maximizing the entropy during the
first few fm/c of the reaction; isotropization could be
sufficient~\cite{Arnold:2004ti}. In fact, data~\cite{phobos} from RHIC
indicate that the number of charged particles per participant {\em in
the final state} is only $\sim 30\%$ lower in central d+Au collisions
than it is in central Au+Au. This perhaps indicates that the
equilibration process expected to occur in Au+Au (but not in d+Au)
does not produce a large amount of entropy~\cite{Pratt}.  Hence, the
mechanism of isotropization of particles via strong fields could be
very interesting for the phenomenology of heavy-ion collisions.

In this letter we solve the classical transport equation
for hard gluons with non-Abelian color charge $Q^a$ in the
collisionless approximation~\cite{ClTransp},
\begin{equation}
 p^{\mu}[\partial_\mu - gQ^aF^a_{\mu\nu}\partial^\nu_p
    - gf_{abc}A^b_\mu Q^c\partial_{Q^a}]f(x,p,Q)=0~.
\end{equation}
It is coupled to the Yang-Mills equation
\begin{equation}
 D_\mu F^{\mu\nu} = j^\nu 
 = g \int \frac{d^3p}{(2\pi)^3} dQ Q v^\nu f(x,p,Q),
\end{equation}
where $f(x,p,Q)$ denotes the one-particle phase space distribution
function~\cite{ClTransp}. These equations were
shown to reproduce the ``hard thermal loop'' effective
action~\cite{ClTransp} near equilibrium. If fluctuations on top of the
mean fields are not neglected, one obtains a collision term from their
moments~\cite{flucs}. 
The same set of transport equations were also derived within the
worldline formalism for the one loop effective action of
QCD~\cite{Jalilian-Marian:1999xt}; the emergence of classical
transport from a quantum kinetic equation derived within the
closed-time-path formalism was discussed in ref.~\cite{WongRedlich}.
For recent reviews of semi-classical transport
theory for non-Abelian plasmas see ref.~\cite{review_class_transp}.
Furthermore, we refer to ref.~\cite{abelian} for a study of particle
production and propagation in Abelian fields, including
back-reaction and collisions in the relaxation time approximation. The
specific point of the present letter, however, is to study possible non-Abelian
plasma instabilities due to anisotropic particle
distributions~\cite{Randrup,ALM,HTL,HTL_unst_mode,Arnold:2004ih,Rebhan:2004ur,Arnold:2004ti}.

We employ the test-particle method~\cite{Bertsch:1988ik}, replacing
the continuous distribution $f(\bm{x},\bm{p},Q)$
by a large number of test particles:
\begin{equation}
 f(\bm{x},\bm{p},Q) = \frac{1}{N_{test}}\sum_i 
  \delta(\bm{x}-\bm{x}_i) (2\pi)^3 \delta(\bm{p}-\bm{p}_i)\delta(Q_i-Q),
\end{equation}
where $\bm{r}_i$ and $\bm{p}_i$ are the coordinates of an individual test
particle. This {\sl Ansatz} leads to Wong's equations~\cite{ClTransp,Wong}
\begin{eqnarray}
\frac{d\bm{x}_i}{dt} &=& \bm{v}_i,\\
\frac{d\bm{p}_i}{dt} &=& gQ_i^a \left( \bm{E}^a + \bm{v}_i \times
  \bm{B}^a \right),\label{pdot}\\
\frac{dQ_i}{dt} &=& igv^{\mu}_i [ A_\mu, Q_i]
\end{eqnarray}
for the $i$-th test particle~\cite{scaleg2}.

The time evolution of the Yang-Mills field can be followed
by the standard Hamiltonian method~\cite{Ambjorn:1990pu}.
Numerical techniques to solve the classical field equations coupled to
particles have been developed in ref.~\cite{HuMullerMoore}.
Our update algorithm is closely related to the one
explained there which generalizes the Abelian version of 
the charge conservation method in particle simulations~\cite{VB1992}.

In the following, we assume that the fields only depend on time and
on one spatial coordinate, $x$, which reduces the Yang-Mills equations to
1+1 dimensions. The hard modes represented by classical particles are
allowed to propagate in three spatial dimensions.
For simplicity, we also restrict ourselves
to the case without expansion here; the more realistic case with
longitudinal expansion~\cite{KV} will be addressed in the future.

The initial anisotropic phase-space distribution of hard gluons is taken to be
\begin{equation}
  f(\bm{p},\bm{x})\propto e^{-\sqrt{p_y^2+p_z^2}/p_\mathrm{hard}}\delta(p_x)~.
\end{equation}
This represents a quasi-thermal distribution in two dimensions, with
``temperature'' $p_\mathrm{hard}$ which now takes over the role of the
saturation momentum mentioned above. We have checked explicitly that
no instability occurs when the particle distribution is taken to be isotropic.

The initial field amplitudes are sampled from a Gaussian distribution
with a width tuned to a given initial energy density.
We solve the Yang-Mills equations in $A^0=0$ gauge and also set
$\bm{A}=0$ (i.e.\ all gauge links =1) at time $t=0$; the initial electric
field is taken to be polarized in a random direction transverse to the x-axis.
Gauss' law is then used to obtain the initial charge distribution. 
All results shown below were obtained using a lattice with $N=512$ sites;
we have checked the numerical accuracy by comparing to
$N=256$, 1024  lattices
(for the same physical parameters) and by monitoring conservation of the total
energy and of Gauss' law. The total energy was conserved to within
$5\times10^{-4}$ ($5\times10^{-3}$)
over the course of the simulations for the weak
(strong) field initial conditions, and the maximal violation
of Gauss' law was $10^{-9}$ for $SU(2)$ and $10^{-33}$ for $U(1)$ (in
lattice units).

Before coming to our results, we also comment on the occurence of
``anomalous Cherenkov radiation''. This corresponds to anomalous hard
radiation from soft modes which may occur for simulations on a
discrete lattice, as the dispersion relation of the fields may contain real
space-like modes. For example, taking $\omega(k)=k$ for free fields
in the continuum
leads to the dispersion relation $\omega(k)=2|\sin(ka/2)|/a$ on a
one-dimensional lattice ($|k|<\pi/a$).
Consequently, hard field modes with $k\sim1/a$
would then get populated on the lattice because their ``mass''
$\sqrt{\omega^2 -k^2}$ is imaginary. The situation could perhaps be
improved by employing higher-order discretization schemes for the Yang-Mills
action or by damping hard modes exponentially at $t=0$. 
However, we have not done so at present. Our tests with
different lattices do not indicate a significant
dependence of either the growth or the saturation of the instability on
the lattice spacing. Also, our solutions for isotropic particle
momentum distributions appear to be stable when the number of test
particles is taken to infinity.
A possible physical reason for this observation
could be that interactions among soft modes with $k\ll 1/a$ and with large
occupation numbers, and interactions of those modes with the
particles, which give the largest contribution to the total energy
(see below), actually dominate.  Nevertheless, this effect may deserve
a more careful numerical study in the future.

\begin{figure}[ht]
\includegraphics[width=3.3in]{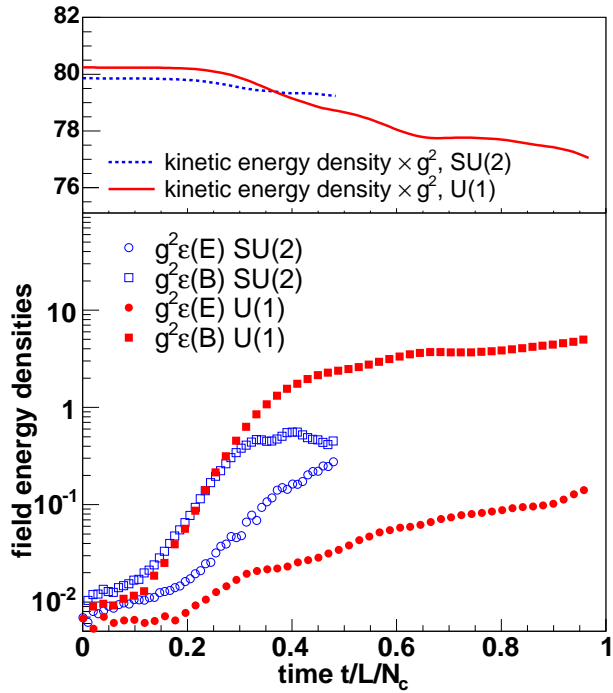}
\caption{
Time evolution of the kinetic (particle), magnetic and electric energy
densities in GeV/fm$^3$ for $U(1)$ and $SU(2)$ gauge group, respectively.
}
\label{fig:fieldW}
\end{figure}

We first show results for a large separation of initial particle and field
energy densities which should qualitatively resemble the conditions
studied in~\cite{Arnold:2004ih,Rebhan:2004ur,Arnold:2004ti}. 
The results shown in
Fig.~\ref{fig:fieldW} corresponds to a lattice of physical size
$L=40$~fm, a hard scale  $p_\mathrm{hard}=10$~GeV and a particle
density of {$n/g^2= 10$~fm$^{-3}$.}

For the $U(1)$ case we observe a rapid exponential growth of the
magnetic field energy density starting at about $t/L\approx 0.1$,
turning into a slower growth at $t/L\approx 0.5$; at this point the
magnetic fields have grown sufficiently to affect the particles which
visibly start loosing energy.  The electric field grows less rapidly
and equipartitioning is not achieved within the depicted time
interval. This indicates that the
field strengths are still too high for linear response to apply.
In the non-Abelian case the growth of the magnetic field
saturates earlier, and the electric field has comparable strength by
the end of the simulation. Also, it appears that the saturation of the
magnetic instability occurs before it has a noticeable effect on the
particles since their energy density is nearly constant. Nevertheless,
at a purely qualitative level the $U(1)$ and $SU(2)$ simulations are
not vastly different, as anticipated in~\cite{Arnold:2004ih}.

\begin{figure}[ht]
\includegraphics[width=3.3in]{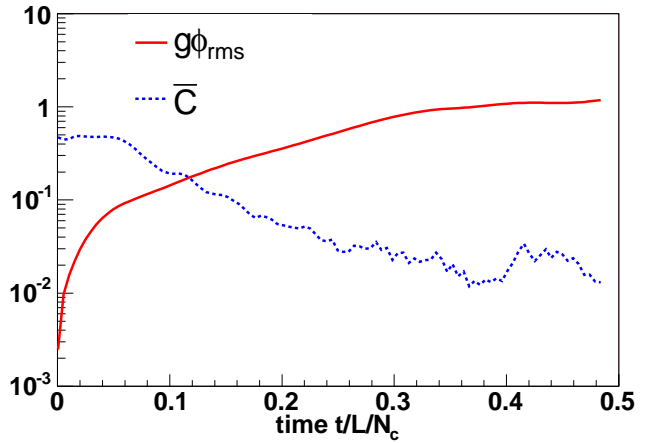}
\caption{
The average amplitude $\phi_\mathrm{rms}$ (in units of GeV)
and the relative size
$\bar{C}$ of commutators as a function of time; physical parameters as
in Fig.~\ref{fig:fieldW}.
}
\label{fig:RMSCbarW}
\end{figure}
This is analyzed further in Fig.~\ref{fig:RMSCbarW},
showing the growth of the rms average
\begin{equation}
  \phi_\mathrm{rms}= \left[
    \int_0^L \frac{dx}{L} (A^a_yA^a_y + A^a_zA^a_z)\right]^{1/2}~,
\end{equation}
and the average of the relative size of the field commutator defined
by~\cite{Arnold:2004ih}
\begin{equation}
  \bar{C} = \int_0^L \frac{dx}{L} \frac{
         \sqrt{\mathrm{Tr}\; ( (i[A_y,A_z])^2 )}
	 }{\mathrm{Tr}\; (A_y^2 + A_z^2)}~.
\end{equation}
The behavior of $\phi_\mathrm{rms}$ is similar to that of the field
energy density shown above. Initially, $\bar{C}$ is constant but then 
starts dropping exponentially when the magnetic instability sets in,
indicating the partial ``abelianization'' of the
fields~\cite{Arnold:2004ih,Rebhan:2004ur}. The rate by which $\bar{C}$
drops in the intermediate stage is
roughly comparable to the growth rate of $\phi_\mathrm{rms}$; also,
the abelianization appears to stop after $\bar{C}$ dropped by about
one order of magnitude, at about the same time when the exponential
growth of the fields and of $\phi_\mathrm{rms}$ saturates.

\begin{figure}[ht]
\includegraphics[width=3.4in]{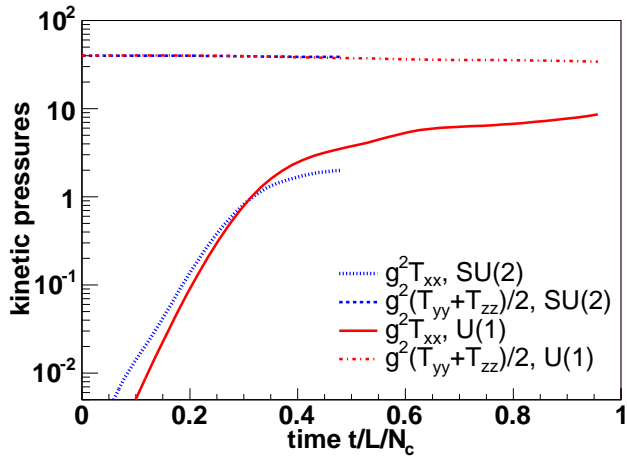}
\caption{Transverse and longitudinal components of the energy-momentum
  tensor of the particles for the simulation corresponding to
  Figs.~\ref{fig:fieldW},\ref{fig:RMSCbarW} (weak field).
}
\label{fig:kineticW}
\end{figure}
Finally, in Fig.~\ref{fig:kineticW} we show the time evolution of the
longitudinal and transverse components of the energy-momentum tensor
of the particles, i.e.\ the kinetic pressure. For both $U(1)$ and
$SU(2)$ we observe a rapid growth of the longitudinal pressure, which
is zero initially. Again, the rate is somewhat smaller for the
non-Abelian case. The approach to ``isotropization'' of the kinetic
pressure is clearly correlated to the stage of exponential growth of
the soft fields seen before~\cite{Arnold:2004ti}. However, for both
cases $T_{xx}$ remains significantly smaller than the transverse
component for times $\lsim L$.

The initial conditions above were chosen such as to verify
qualitatively the picture emerging in the hard loop approximation, where
the field energy density is (and remains) much smaller than that of
the particles and so the back-reaction can be
neglected~\cite{Arnold:2004ih,Rebhan:2004ur,Arnold:2004ti}. In the color glass
condensate model of high-energy collisions one does not expect such a
strong separation of energy densities, however. Since our numerical
solution includes the back-reaction of the fields on the particles, we
study the situation with stronger fields next.

Specifically, the simulations below were performed with the following
set of physical parameters: length $L=10$~fm, hard scale
$p_\mathrm{hard}=1$ GeV, particle density {$n/g^2= 500$~fm$^{-3}$} and an
initial field energy density of about 20~GeV/fm$^3$.

\begin{figure}[ht]
\includegraphics[width=3.3in]{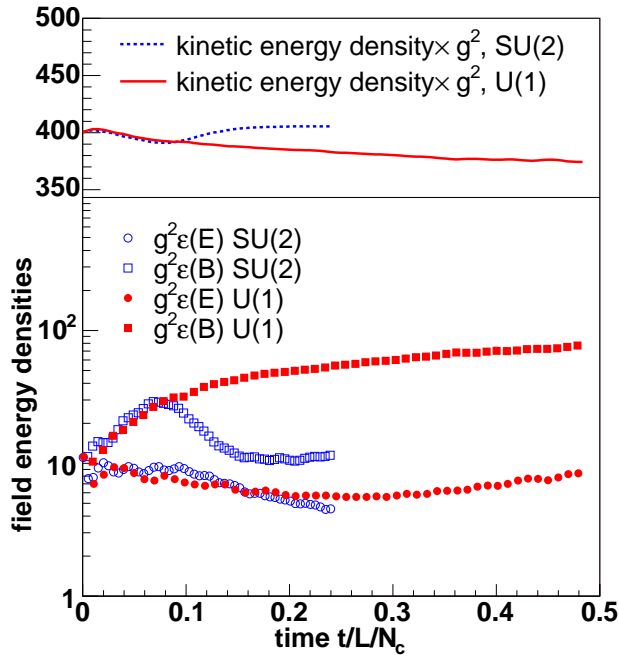}
\caption{
Time evolution of the kinetic and field energy densities for strong
initial fields.
}
\label{fig:fieldS}
\end{figure}
Fig.~\ref{fig:fieldS} shows the time evolution of the energy densities
for these initial conditions. This case clearly differs from the
weak-field limit shown before. Over the time interval shown, the
electric field energy density is practically constant for both $U(1)$
and $SU(2)$. The Abelian magnetic field does exhibit a slow growth,
draining some energy from the particle reservoir. For $SU(2)$,
however, after a short initial growth the magnetic field energy
decreases again, to saturate pretty much at its initial value. 
Therefore, the particle energy density is also more or less constant
over the depicted time interval.

\begin{figure}[ht]
\includegraphics[width=3.3in]{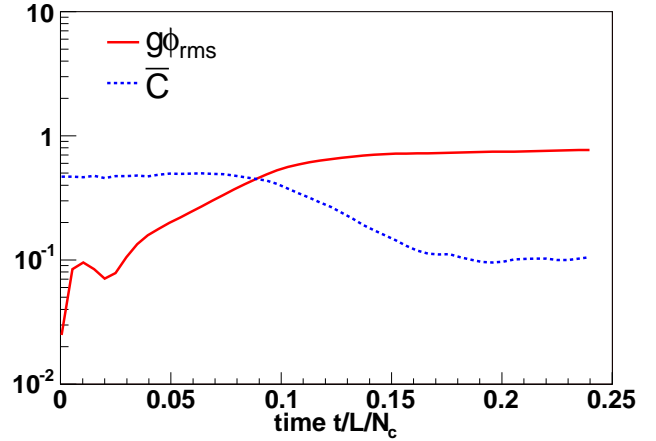}
\caption{Time evolution of $\phi_\mathrm{rms}$ and $\bar{C}$ in the
  strong field case.
}
\label{fig:RMSCbarS}
\end{figure}
Fig.~\ref{fig:RMSCbarS} confirms this observation via the
$\phi_\mathrm{rms}$ observable: the initial growth saturates much
earlier than before. Similarly, the average
commutator $\bar{C}$ stays constant for some time then drops by about
a factor 5 (during the period where the magentic field drops~!) and
saturates at $\approx 10\%$, which is an order of
magnitude larger than for the weak field case from
Fig.~\ref{fig:RMSCbarW}. This indicates a much smaller degree of
``abelianization'' for strong fields.

\begin{figure}[ht]
\includegraphics[width=3.4in]{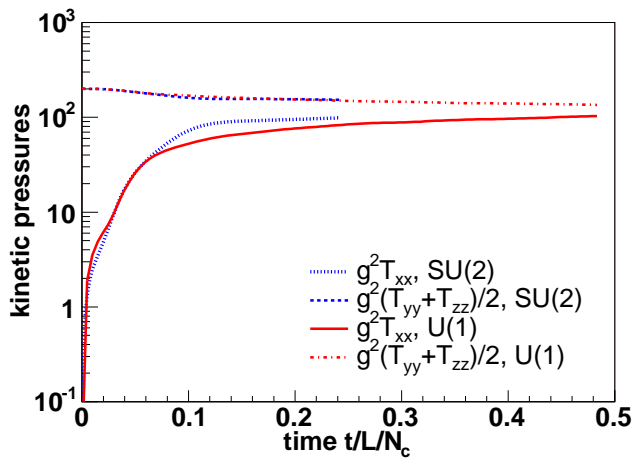}
\caption{
Transverse and longitudinal components of the energy-momentum
  tensor of the particles for the simulation corresponding to
  Figs.~\ref{fig:fieldS},\ref{fig:RMSCbarS} (strong field).
}
\label{fig:kineticS}
\end{figure}
Perhaps surprisingly, Fig.~\ref{fig:kineticS} nevertheless shows a
very rapid isotropization of the kinetic pressure for both $U(1)$ and
$SU(2)$ (note that $t/L=0.1$ corresponds to t=1~fm in physical units
for this lattice). Moreover, the {\em degree} of isotropization is
much higher, i.e.\ the transverse and longitudinal pressures are
closer than in Fig.~\ref{fig:kineticW}. The very fast and nearly
complete isotropization is, of course, the reason why field
instabilities can not be sustained over a significant period of time
in this case. It is caused by the bending of the particle
trajectories in the strong field which is very different
from conventional parton cascade transport with small-angle
perturbative scattering (and no field). The random initial
fields then cause a rapid isotropization of the particle momenta via
eq.~(\ref{pdot}). Note that this does not require hard modes in the
fields, which indeed would violate the assumed separation of momentum
scales, but large field amplitudes.

In summary, we have studied instabilities in the
coupled Wong Yang-Mills equations for strongly anisotropic initial
particle momentum distributions. For both $U(1)$ and $SU(2)$ gauge
groups we do observe a period of exponential growth of the fields when
their initial energy density is far less than that of the hard modes
(particles). This, in turn, leads to partial isotropization of the
particle momentum distributions and of the kinetic pressure. Although we
find somewhat smaller field growth and isotropization rates for the
non-Abelian case, we nevertheless qualitatively confirm the picture
developed in~\cite{Arnold:2004ih,Rebhan:2004ur} in that the
non-Abelian fields ``abelianize'' efficiently
during the period of exponential growth.

For large initial field amplitudes, corresponding to a smaller ratio
of initial particle to field energy densities, our results are
qualitatively different. We observe a very rapid isotropization of the
particle momentum distributions which is due to bending of their
trajectories in the strong fields on a time-scale that is
relevant for the physics of high-energy collisions. This, however,
inhibits the development of instabilities of the Yang-Mills
fields. Nevertheless, these results, too, suggest that the presence of
the strong non-Abelian fields should be taken into account to
understand the process of isotropization in the early stages of
high-energy collisions.

\begin{acknowledgments}
We thank F.~Gelis, M.~Gyulassy, A.~Mueller, R.~Pisarski and M.~Strickland
for helpful discussions and J.~Lenaghan and
M.~Strickland for drawing our attention to this subject.
\end{acknowledgments}

\vspace*{.2cm}
{\bf Note added:} After this manuscript was submitted for publication,
a paper appeared~\cite{MM} which presents an analytical discussion of
a possible effective potential for anisotropic QCD plasmas beyond the hard loop
approximation. Also, a modified ``bottom-up'' scenario for gluon
thermalization in high-energy heavy-ion collisions~\cite{MSW} and
$3+1d$ simulations of the full hard-loop effective theory~\cite{AMY} appeared.


\begin{figure*}[htb]
\includegraphics[width=3.4in]{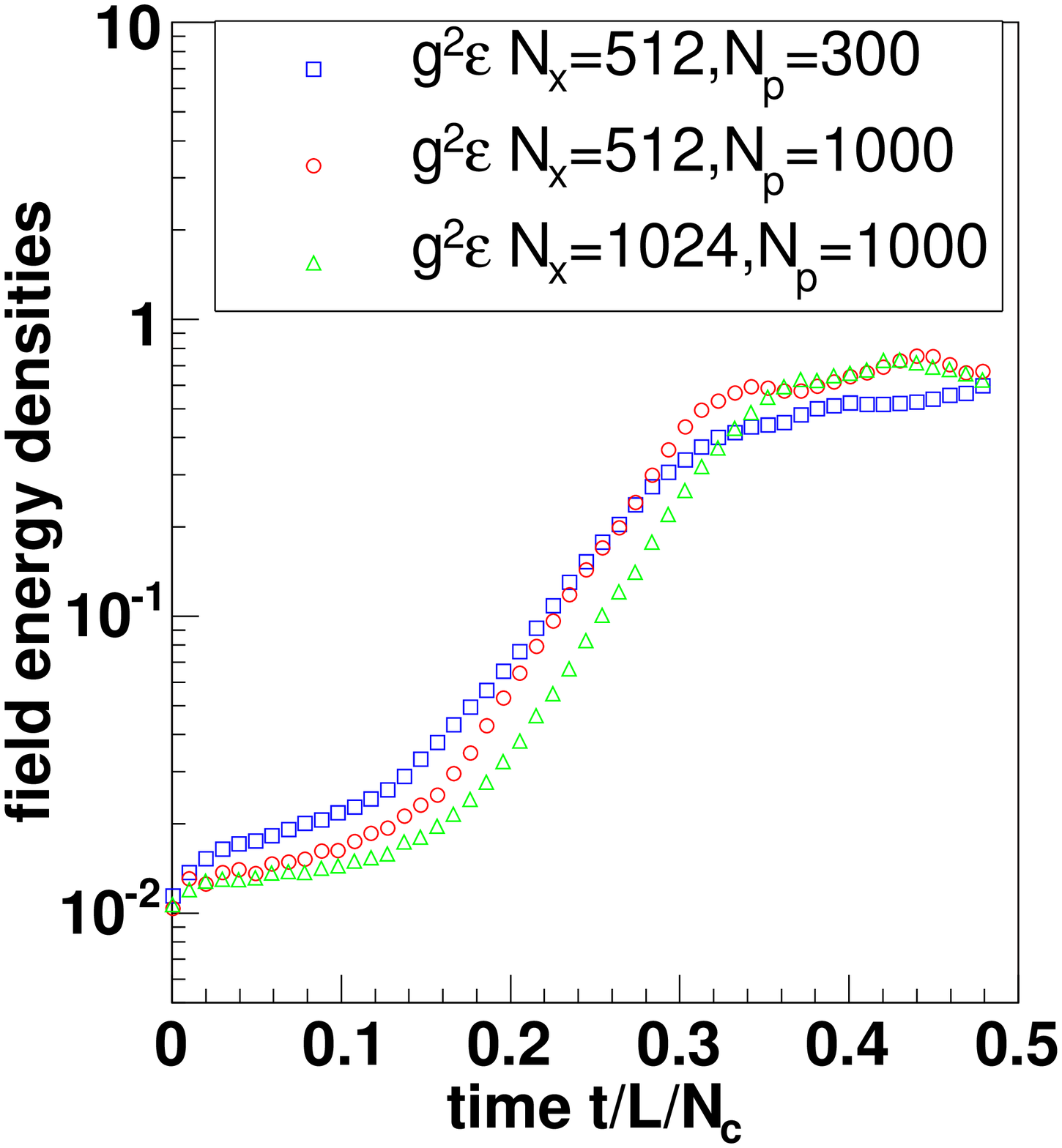}
\caption{
ADDITIONAL FIGURE: Time evolution of the field energy density on
two different lattices (with the same
physical size $L$); $N_p$ denotes the number of test particles
per lattice site.
}
\label{add1}
\end{figure*}

\begin{figure*}[htb]
\includegraphics[width=3.4in]{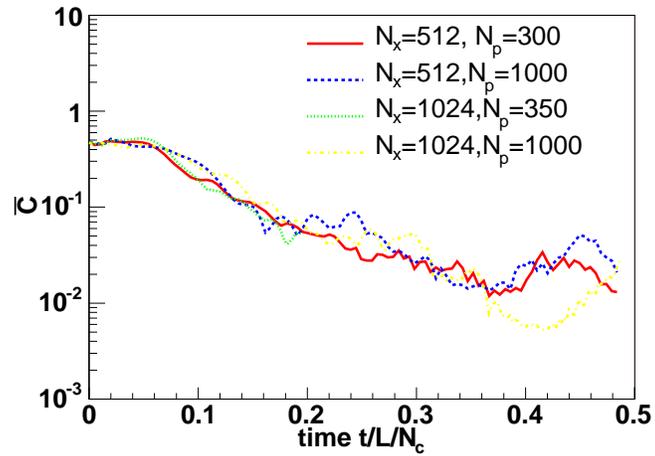}
\caption{
ADDITIONAL FIGURE: Same for the commutator.
}
\label{add2}
\end{figure*}

\begin{figure*}[htb]
\includegraphics[width=3.4in]{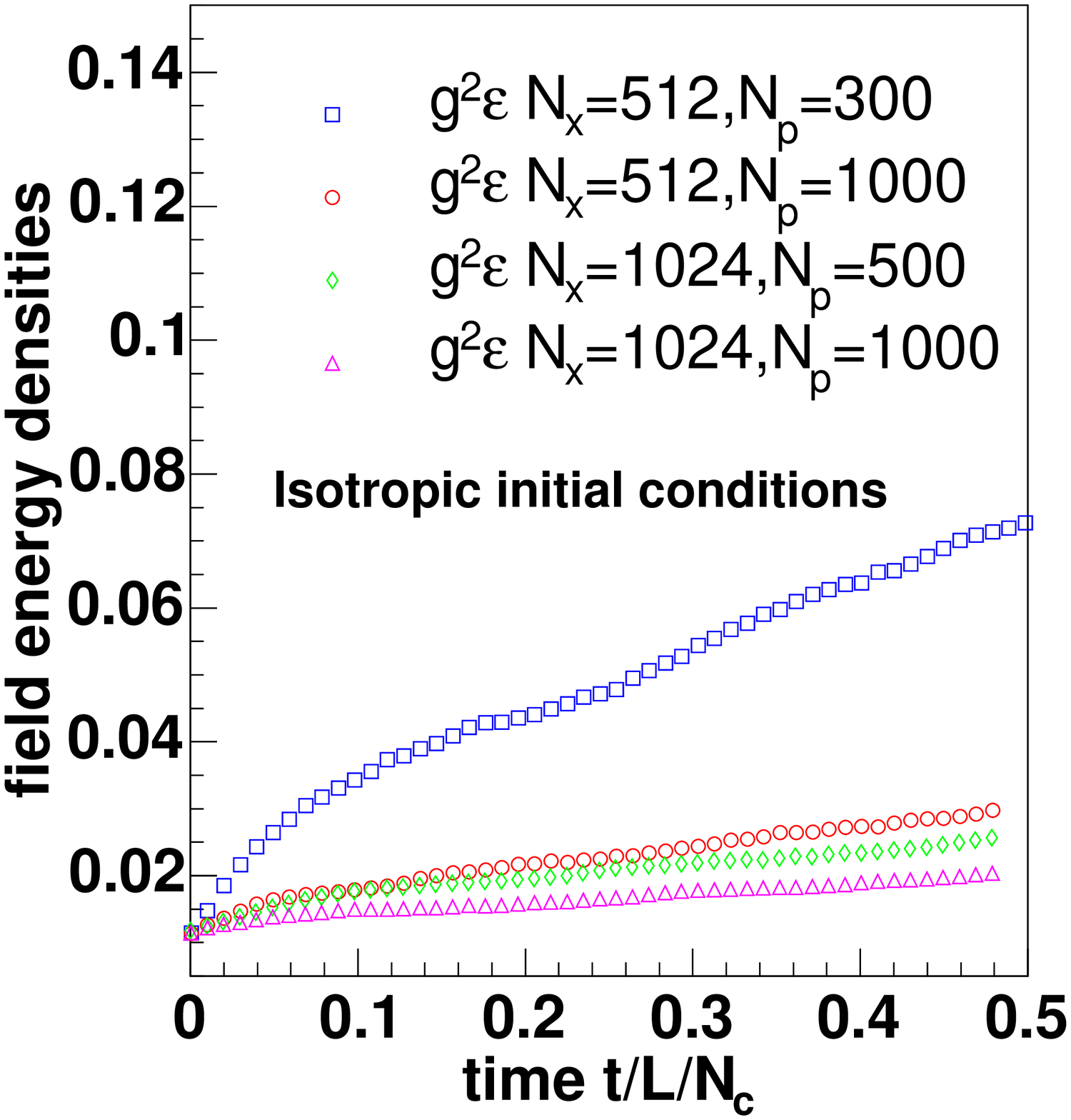}
\caption{
ADDITIONAL FIGURE: time evolution of the field energy density for {\em
  isotropic} particle momentum distributions.
}
\label{add3}
\end{figure*}


\begin{thebibliography}{99}
\bibitem{scaleg2}
In our numerical calculations
we redefine the Hamiltonian $H'=g^2 H$, the fields
$A_\mu' = g A_\mu$ and the particle phase space density $f'=g^2 f$
in order to remove the gauge coupling from the classical theory.
\end{thebibliography}
\end{document}